# THE MEASUREMENT OF BOVINE PERICARDIUM DENSITY AND ITS IMPLICATIONS ON LEAFLET STRESS DISTRIBUTION IN BIOPROSTHETIC HEART VALVES

Masod Sadipour and Ali N. Azadani

Department of Mechanical and Materials Engineering

University of Denver, Denver, CO, USA

**Corresponding author:**

Ali N. Azadani, Ph.D.

Associate Professor

University of Denver

Department of Mechanical and Materials Engineering

2155 E. Wesley Ave #439

Denver, CO 80208

Phone: 303-871-3636

Fax: 303-871-4450

Email: ali.azadani@du.edu

## ABSTRACT

**Purpose**: Bioprosthetic Heart Valves (BHVs) are currently in widespread use with promising outcomes. Computational modeling provides a framework for quantitatively describing BHVs in the preclinical phase. To obtain reliable solutions in computational modeling, it is essential to consider accurate leaflet properties such as mechanical properties and density. Bovine pericardium (BP) is widely used as BHV leaflets. Previous computational studies assume BP density to be close to the density of water or blood. However, BP leaflets undergo multiple treatments such as fixation and anti-calcification. The present study aims to measure the density of the BP used in BHVs and determine its effect on leaflet stress distribution.

**Methods:** We determined the density of eight square BP samples laser cut from Edwards BP patches. The weight of specimens was measured using an A&D Analytical Balance, and volume was measured by high-resolution imaging. Finite element models of a BHV similar to PERIMOUNT Magna were developed in ABAQUS.

**Results:** The average density value of the BP samples was 1410 kg/m$^3$. In the acceleration phase of a cardiac cycle, the maximum stress value reached 1.89 MPa for a density value of 1410 kg/m$^3$, and 2.47 MPa for a density of 1000 kg/m$^3$ (30.7% difference). In the deceleration, the maximum stress value reached 713 and 669 kPa, respectively.

**Conclusion:** Stress distribution and deformation of BHV leaflets are dependent upon the magnitude of density. Ascertaining an accurate value for the density of BHV leaflets is essential for computational models.

## 1. INTRODUCTION

Bioprosthetic heart valves (BHVs) are a non-invasive promising treatment option for people suffering from heart valve disease. BHVs do not demand long-term anticoagulant therapy while they have better hemodynamics as compared to mechanical heart valves [1–2]. Bovine pericardium (BP) is one of the common materials in manufacturing BHVs which includes a high amount of layered proteins (collagen and elastin) [1−4]. In BHV implants fabrication, picked BP patches are cut into leaflet shapes to apply onto sewing rings to form BHVs. It has been shown that valve performance is highly dependent on leaflet material properties. Considering and analyzing these characteristics is of interest to many researchers. In 1998, Hiester et al. [5] examined the optimal sections of BP for use in BHV by investigating 20 BP sacs. An optimal BP site for BHV needs to have low-fiber preferred direction variation across the site and negligible amounts of fat. Hiester et al. used small-angle-light scattering (SALS) to report the fiber architecture and preferred direction distribution. They found some regions in all BP sacs that meet criteria. In addition, they indicated that each BP sac has considerable variation in the location and fiber orientation of the optimal sites. Structural valve degeneration is the primary cause of failure in bioprosthetic valves. Procedures to prevent SVD in pericardial valves have concentrated on two main areas. First, developing anti-calcification treatments to prevent or slow leaflet calcification; and second, modifying the method of leaflet suspension to minimize mechanical stress. Nearly all current commercially available bioprosthetic valves utilize glutaraldehyde (GA) fixation in their fabrication. GA fixation results in cross-linking of collagen and antigen (chemical stabilization) reduction, and also reduces immunogenicity when implanted. GA fixation at high-pressure distorts crimp geometry and matrix fragmentation—causing stiffening and kinks in valve leaflets. Thus, fixation is usually done under physiological pressure to minimize these problems[6][7]. The GA fixation process leaves residual free aldehyde components and phospholipids, which can act as binding sites for calcium, leading to early, severe calcification and structural deterioration if left untreated [8]. In 2011, Gilberto et al.[6] worked on other chemical approaches to prepare an acellular bovine pericardium matrix for BHVs production. They utilized an alkaline extraction in calcium salts for cell removal. Gilberto et al. observed that while they had cell removal after 12 hours, collagen fibril structure and stability to collagenase increased after 24 hours of the process. Their results revealed a process for manufacturing tissue valves and tissue reconstruction. In 2017, Tam et al. [3] investigated the effect of replacing Glutaraldehyde with an irreversible, carbodiimide-based crosslinking chemistry (TRI) to increase valve performance regarding calcification and tissue degeneration. They observed that TRI curb calcification like other solutions, while providing more robust tissue through stabilizing more of the tissue ECM. The tissue is more compliant and resists permanent deformation and degeneration. Likewise, in 2018, Meuris et al. [9] introduced a novel treatment for reducing the mineralization of BPHV to increase tissue durability by using an aldehyde-free solution for tissue storage (free-treated precordium). They compared mechanical and biomechanical characteristics of the treated valve with samples from commercially available biological values (Trifecta and Perimount Magna). They observed that the novel treated pericardium's mechanical strength and biomechanical properties are related to mentioned commercial valves. They revealed that mitral implants of novel-treated valves in sheep had great hemodynamic performance at five months without any calcification. They concluded their treatment enhances the anticalcification properties and may thereby improve long-term durability of the valve.

The widely recognized Perimount bovine pericardial valves manufactured by Edwards Lifesciences originally used the XenoLogiX tissue treatment, involving a phospholipid removal from the leaflets through a two-steps process [10]. It was shown by Cunanan et al. [10] that the aforementioned treatment proved effective in removing above 90% of the phospholipids from the leaflets; notably, these leaflets serve as calcification binding sites. The anti-calcification procedures have been growing and improving during the past three decades at Edwards. In order to fix, sterilize, and decrease pericardial leaflet antigenicity, buffered glutaraldehyde and FET (formaldehyde-tween 80 solution) are employed by the TFX (Thermafix) method [11]. The Perimount Magna Ease tissue valve has used the same process. A novel treatment has been offered by Edwards recently, based upon perpetual blocking of all interactions with calcium by capping the free-aldehydes. Subsequently, glycerolisation occurs on the tissue with the purpose of substituting any remaining molecules of water with glycerol. A 72% decrease in calcium was achieved through application of this tissue (a.k.a., the "Resilia" tissue) in a Perimount valve, relative to a valve in a juvenile sheep model treated via TFX [12]. In the period of 2013–16, 689 patients received Perimount Magna Ease valve implants through the COMMENCE trial, which were reformed to entail the Resilia tissue [13]. Data over the subsequent four years demonstrated desirable hemodynamic performance and safety (with mean gradients of $11.0 \pm 5.6$ mmHg, effective orifice areas of $1.5 \pm 0.5$ cm$^2$, and no cases of SVD [14]).

On the other hand, other studies investigated the mechanical properties of different sections of bovine pericardium. In 1998, Hiester et al.[5][15] examined 20 BP sacs to locate optimal BP sections for use in BHV. An optimal BP site for BHV needs to have low-fiber preferred direction variation across the site and negligible amounts of fat. They used small-angle-light scattering (SALS) to report fiber architecture and preferred direction distribution. They found some regions in all BP sacs that meet the criteria. They also noted that each BP sac has considerable variation in location and fiber orientation in the optimal sites.

In vitro studies play a significant role in describing the mechanical properties of soft tissues as well as computational modeling validation. Finite element analysis (FEA) has been widely used to study prosthetic and natural heart valves, reducing the costly and lengthy process required for in vivo and in vitro analysis [4]. Finite element (FE) simulation is a powerful tool in the analysis of heart valve durability. In numerical simulations of BHVs, some challenges include leaflet contact and nonlinear anisotropic leaflet mechanical properties. Experimental tests exist to validate results [14, 16]. Utilization of actual leaflet material properties—such as mechanical properties, damping coefficient, and density—is essential for accurate BHV FE simulations. Bioprosthetic leaflet mechanical properties are shown to be mechanically anisotropic [17]. Different numerical simulations of heart valves showed that the anisotropic Fung model is a reliable model in BHVs simulations that can accurately predict valve deformation during opening and closing of the valve. Abassi et al. [18] utilized inverse FE methods to determine anisotropic material constants of bovine pericardial leaflets of a surgical bioprosthetic heart valve. Like other BHV mechanical properties (Isotropic and anisotropic mechanical models), BHV density has a considerable impact on leaflet stress and strain distribution. Previous computational studies assumed BP density to be close to the density of water or blood (1000–1100 kg/m$^3$). However, BP leaflets undergo multiple treatments as described earlier. In 2013, Borazjani [19] studied a Fluid-Structure Interaction (FSI) simulation of bio-prosthetic heart valves by using immersed boundary Navier-Stokes (NS) equations for the fluid domain and a large deformation FE method for the structure. In this study, three shells with a density of 1060 kg/m$^3$ and a thickness of 1 mm were considered for the leaflets. They reported a considerable difference between the blood flow field passing through the tissue valve and mechanical valve during the heartbeat cycle. Small-scale vortical structures were observed in the mechanical heart valve before the peak of systole, but not in the bio-prosthetic heart valves. The results of incomplete transcatheter aortic valve (TAV) expansion on the leaflets stress and strain distribution was considered using FEA [20]. In this study, 1100 kg/m$^3$ was considered the leaflet's density. It was observed that different sizes of incomplete expansion of TAV stent can make localized high-stress regions within the commissure and belly of TAV leaflets. Another study considered the effect of geometry on leaflet stresses in Bicuspid aortic valves (BAV) using FEA[21]. Material properties obtained from porcine aortic valve leaflets was employed for the simulation. A density of 1100 kg/m$^3$ and thickness of 0.386 mm was set in Abaqus FEA software. Significant changes in leaflet strain and stress were observed with variations in the geometry of simulated BAV. They concluded that calcific aortic stenosis, which frequently exists in patients with BAV, may result from geometrical variations.

The goal of the present work is to measure the density of bovine pericardial patch and analyze the impact of density on BHV leaflet stress and strain distribution. To do so, we conducted an FE analysis of surgical valves by using the anisotropic Fung model and measured density. Finally, we optimized the material coefficients of the Fung model with the new density. To the best of our knowledge, this is the first study to measure the density of commercially available bovine heart valves, and compare leaflet deformation and stress with different densities.

## 2. METHODS
*Experimental Testing*
The density of Edwards bovine pericardial patch (Model 4700, Edwards Life Sciences, CA) used in BHVs was directly measured in the present study. Accordingly, eight specimens were cut from two bovine pericardium patches (four from each patch). A Fusion M2 40 (Epilog Laser, CO) laser cutter was employed to excise square-shaped specimens (10×10 mm), and patches were stored in a saline container. A tissue thickness value of 0.5 mm ± 0.25 mm is reported on the Edwards Lifesciences website [22]. A dial caliper (Fowler 52-008-712-0, Canton, MA) was used to measure all edges of the specimens. In addition, pictures were taken from the specimens on a piece of Teflon and were processed using Photoshop to more accurately determine sample length and thickness. Specimen dimensions were measured with ±0.01 mm accuracy.

To measure leaflet density as blood passes through, we put the leaflets in a solution with the human blood density. Accordingly, water was mixed with glycerol 99% to produce a solution with blood density (1.056 g/ml) at normal human body temperature (37°C). The volume ratio was 9.0 ml for glycerol and 31.0 ml for the water, as previously reported by Glycerine Producers' Association[23].

We considered two procedures for measuring the weight of the samples. In the first one, sample weight was measured inside of the prepared solution. Consequently, specimens were placed in the solution with blood density to measure the weight after removing the container droplets. In the second procedure, sample weight was measured by directly placing samples on the Analytical Balance (A&D GR-202, Gurgaon, Haryana). The Analytical Balance used in this study had an accuracy of $10^{-4}$ g.

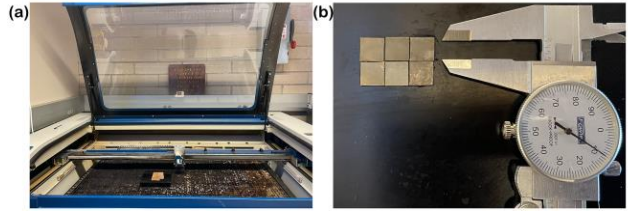

**Figure 1. a) Fusion M2 40 laser cutter and samples b) Samples thickness measuring using dial caliper.**

*Computational Modeling*
An FE analysis was conducted on the effect of density deviation on the deformation and stress distribution of PERIMOUNT Magna aortic heart valve (Edwards Life Sciences, CA)—a well-known and commercially available surgical valve. For this purpose, we conduted a dynamic analysis on the 25 mm PERIMOUNT Magna Bioprosthesis with five different densities—1,000, 1,100, 1,200, 1,300 and 1,410 kg/m³—based on our finding in the first part of the study. A NextEngine 3D Laser Scanner was also employed to obtain the geometry of the surgical valve leaflets[24] [3]. Meshing was done using Hypermesh. The valve leaflets and the wire frame stent were discretized via approximately 7,700 quad elements (S4) and 400 beam elements, respectively (Fig. 2). Mesh independency was done through three different element numbers.

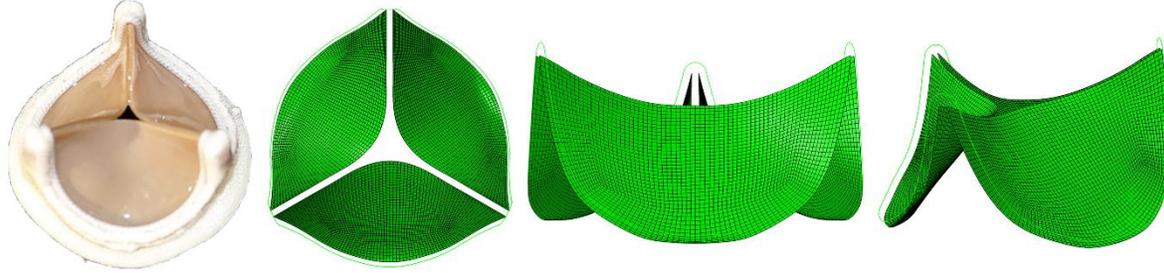

**Figure 2. (Left) 25-mm PERIMOUNT Magna surgical bioprosthetic aortic valve and the FE model showing the leaflets geometric configurations attached to the stent**

Abaqus/Explicit was used to carry out the FE analysis. Material orientation was given to the leaflet's elements utilizing a MATLAB code[24] [24]. A thickness of 0.56 mm was taken into account in the simulations. According to the literature, viscous damping effect of the fluid around the BHV should be considered in the FE analysis via Rayleigh damping[25], which helps achieve a smooth non-oscillatory convergence of the finite element method. Therefore, the Rayleigh damping coefficient was considered as α=10642 1/s in this study[24]. A non-linear anisotropic Fung-type elastic constitutive model was employed for the leaflets:

$w = \frac{c}{2}(e^Q - 1), \ Q = E : (\flat E),$

where E is the Green–Lagrange strain tensor, and ♭ is a non-dimensional symmetric tensor entailing 21 independent coefficients of the Fung model.

$$\begin{bmatrix} b_{1111} & b_{1122} & b_{1133} & b_{1123} & b_{1113} & b_{1112} \\ & b_{2222} & b_{2233} & b_{2223} & b_{2213} & b_{2212} \\ & & b_{3333} & b_{3323} & b_{3313} & b_{3312} \\ & & & b_{2323} & b_{1323} & b_{1223} \\ & \text{Symmetric} & & & b_{1313} & b_{1213} \\ & & & & & b_{1212} \end{bmatrix}$$

A pressure waveform obtained from a previous study using in vitro analysis (Figure 3) was used as the simulation physiological loading condition[24].

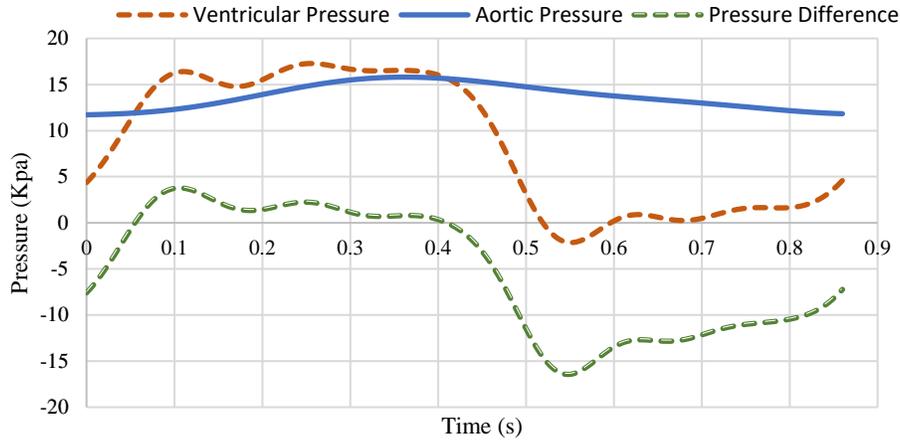

**Figure 3. Pressure wave frame.**

A material optimization was carried out for the leaflets, based on the deformation experimental results of previews study by Mostafa et al. [24]. The experimental hemodynamic results obtained from setting up a 23 mm paramount Magna valve in a pulse duplicator system. A high-speed camera was employed to capture the valves' leaflet motion at 960 FPS. The center of each leaflet was tracked, and the distance between the center of the leaflet and the center of the valve was calculated through the opening and closing the valve. The PSO method from Isight and Abaqus was used in this study as the optimization procedure. Accordingly, the distance between leaflet center and the center of the valve in the FE simulation was compared against the experimental measurements in the pulse duplicator system. The optimization process was set to stop when the objective function's change reduced to less than the tolerance value of $10^{-5}$.

## 3. RESULTS

*Density Measurement*

Square samples of constant size were cut from the bovine pericardium patch and stored in the saline solution. Table 1 shows the length and thickness of each sample from patches 1 and 2. The results of mass measurements are shown in Table 1, as well. As visible, the weight of each sample measured using method 2 is higher than method 1 due to solution absorption in the samples.

**Table 1: Dimensions and volume of specimens; mass of specimens and density of bovine precordium patch**

|  | Case Number | Thickness (mm) | X (mm) | Y (mm) | Volume (mm³) | Weight measured directly(g) | Weight in B Solution(g) | Density: measured directly (g/cm³) | Density: Samples in the solution (g/cm³) |
|---|---|---|---|---|---|---|---|---|---|
| **Patch 1** | 1 | 0.55 | 9.80 | 9.81 | 52.4619 | 0.0681 | 0.0735 | 1.23 | 1.40 |
|  | 2 | 0.57 | 9.78 | 9.85 | 54.9098 | 0.0622 | 0.0783 | 1.13 | 1.43 |
|  | 3 | 0.56 | 9.78 | 9.85 | 53.9464 | 0.0611 | 0.0672 | 1.13 | 1.26 |
|  | 4 | 0.63 | 9.67 | 9.67 | 58.9106 | 0.0759 | 0.0815 | 1.29 | 1.38 |
|  | mean | 0.58 | 9.76 | 9.79 | 55.0572 | 0.0667 | 0.0751 | 1.20 | 1.37 |
|  | SD | 0.04 | 0.06 | 0.08 | 2.7592 | 0.0069 | 0.0062 | 0.068 | 0.064 |
| **Patch 2** | 1 | 0.485 | 9.68 | 9.81 | 46.05599 | 0.0622 | 0.0674 | 1.35 | 1.46 |
| | 2 | 0.49 | 9.70 | 9.8 | 46.5794 | 0.0654 | 0.0699 | 1.40 | 1.50 |
| | 3 | 0.51 | 9.55 | 9.78 | 46.6995 | 0.0660 | 0.0642 | 1.41 | 1.37 |
| | 4 | 0.50 | 9.51 | 9.65 | 45.8375 | 0.0620 | 0.0681 | 1.35 | 1.49 |
| | **mean** | 0.49 | 9.6075 | 9.76 | 46.2931 | 0.0639 | 0.0674 | 1.38 | 1.46 |
| | **SD** | 0.01 | 0.0977 | 0.0744 | 0.4126 | 0.0020 | 0.0023 | 0.027 | 0.056 |

Additionally, the density of each sample and the average for each different patch is shown in Table 1. The standard deviation (SD) for the density for all specimens is less than 0.07, indicating that the values are close to the mean of the set. Finally, by measuring the mean of bovine precordium patch density, in the two different processes, the values 1,300 kg/m$^3$ and 1,410 kg/m$^3$ is reported for the first and second measurement process respectively. As it discussed in the method, the second method in measuring the density is more reliable and 1,410 kg/m$^3$ was used as our reference density in our PERIMOUNT Magna Ease valve simulations and material optimization. As mentioned in the introduction, in previous simulations and analyses, 1000 kg/m$^3$ and 1100 kg/m [19,21,25,26] was used for both bovine and porcine precordium density. Thus, we compared the stress and strain on the leaflets—the most important surgical valve longevity analysis with different densities (1,000, 1,100, 1,200, 1,300, 1,410 kg/m$^3$) in both systole and diastole.

*Leaflet Stress and Strain Distributions*
Fig. 4a and 4b show the maximum value in plane principal stress (MPPS) distributions of the 25 mm premium Magna surgical valve with different density at the acceleration and deceleration. Since dynamic motion exists in the opening and closing of the valve, results were analyzed during acceleration and deceleration. As is visible, the leaflet stress distribution was dependent on the density of the leaflets. Table 2 includes MPPS and the percentage change in the MPPS on the leaflets with 1410 kg/m$^3$ density in comparison with four other densities in acceleration and deceleration. The magnitudes of the MPPS in acceleration are much larger than the corresponding magnitudes in the deceleration. In the acceleration, high-stress regions were detected in the fixed edges between middle of fixed edge and commissure, and the maximum principal stress reached to 1.91 MPa for the valve with the 1410 kg/m$^3$ density. As it is expected, the amount of higher density shows lower stress during the systole. In addition, higher density reduces the acceleration of the leaflets, achievning lower deformation and stress. As presented in Figure 4a, lower density (incorrect density) induced localized high stress and strain regions on the fixed edges of the leaflets in the acceleration. The presence of incorrect density resulted in two local stress concentration spots on the fixed edge site and increased the max in plane stress to 2.47 MPa (more than 30% difference). However, high-stress regions were observed mainly near the commissures during deceleration, and the maximum stress reached to 0.713 MPa. Decrease in density from 1410 kg/m$^3$ to 1000 kg/m$^3$ resulted in decreased magnitudes of the maximum principal stress in the deceleration. (Reason)

As shown in Figure 5, the maximum amount of deformation occurred in the middle of moving edge, the amount of displacement reaches to 6.55 mm during acceleration. By considering 1410 kg/m^3 as a reference value for density, the difference of assuming 1,000, 1,100, 1,200 and 1,300 kg/m$^3$ as density will be remarkable (more than 13%). We have the highest magnitude in case 1 (density = 1,000kg/m$^3$). The difference not only happens in acceleration, but can be seen in deceleration. The max displacement results for different density cases in deceleration are shown in Figure 6. As seen in the figure, the contrary acceleration of the highest displacement is case 5. The computed maximum displacement magnitude and U2 of node one (middle of moving edge) results were then compared for different densities in Table 3. The largest change in the maximum displacement based on the density variations in the BHVs models was 13 %, and 11% in acceleration and deceleration respectively.

Also, these results show that the maximum in-plane stress and displacement occur in case 1 (density=1,000 kg/m$^3$) compared to other cases in acceleration, while it occurs in case 5 (density =1,410 kg/m$^3$) in deceleration.

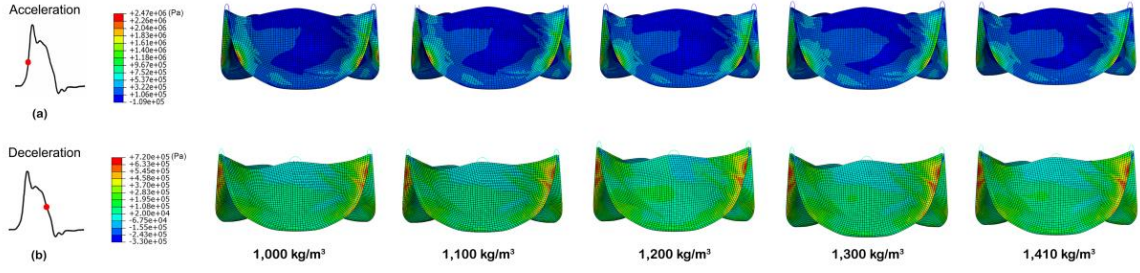

**Figure 4: Max in plane principal stress of 25 mm premium Magna surgical valve in the acceleration and deceleration with different density ranging from 1000 kg/m³ to 1410 kg/m³**

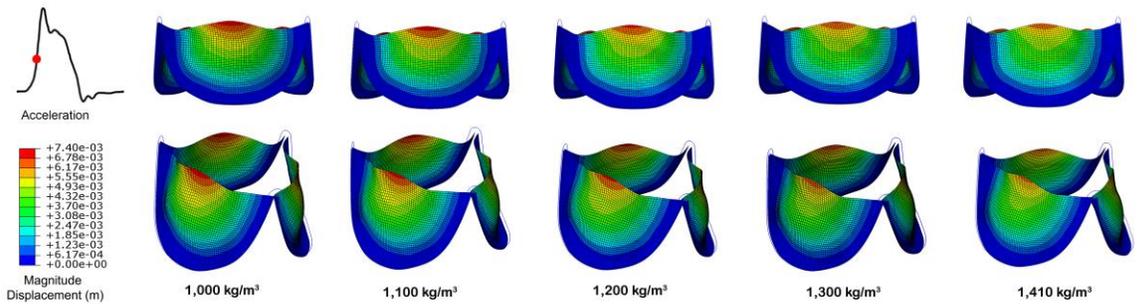

**Figure 5: Displacement magnitude of 25 mm premium Magna surgical valve in the acceleration with different density ranging from 1000 kg/m³ to 1410 kg/m³**

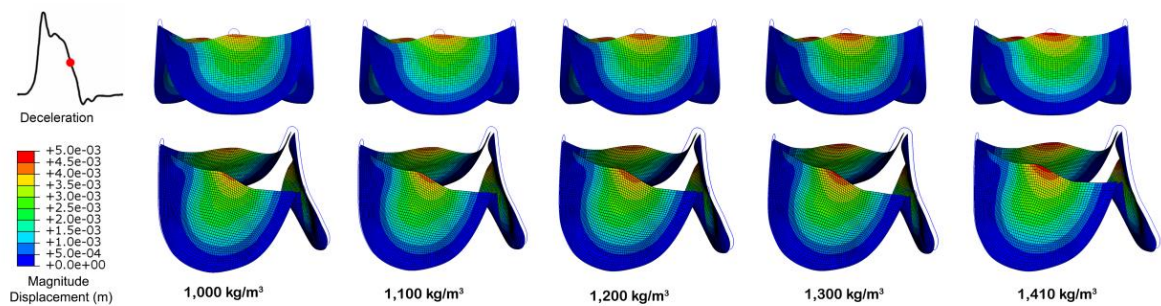

**Figure 6: Displacement magnitude of 25 mm premium Magna surgical valve in the deceleration with different density ranging from 1000 kg/m³ to 1410 kg/m³**

**Table 2: Max in plane stress detected in acceleration and deceleration with different density (ranging from 1000 kg/m³ to 1410 kg/m³)**

| Acceleration | Max in plane stress | Difference (%) | Deceleration | Max in plane stress | Difference (%) |
|---|---|---|---|---|---|
| **Density=1000** | 2.47E+06 | 30.7 | **Density=1000** | 6.69E+05 | 6.17 |
| **Density=1100** | 2.32E+06 | 22.8 | **Density=1100** | 6.81E+05 | 4.48 |
| **Density=1200** | 2.06E+06 | 9 | **Density=1200** | 6.92E+05 | 2.95 |

| | | | | | | |
|---|---|---|---|---|---|---|
| **Density=1300** | 1.91E+06 | 1.06 | **Density=1300** | 7.02E+05 | 1.55 |
| **Density=1410** | 1.89E+06 | 0 | **Density=1410** | 7.13E+05 | 0 |

**Table 3: Magnitude and Node 1 displacement in the acceleration and deceleration with different density ranging from 1000 kg/m³ to 1410 kg/m³**

| Acceleration | U mag. (mm) | Difference (%) | U2 of node 1 | Difference (%) | Deceleration U mag. (mm) | Difference (%) | U2 of node 1 | Difference (%) |
|---|---|---|---|---|---|---|---|---|
| **Density=1000** | 7.4 | 13 | 6.72 | 14.67 | 4.27 | 11.05 | 3.61 | 13.22 |
| **Density=1100** | 7.21 | 10 | 6.54 | 11.6 | 4.42 | 7.92 | 3.76 | 9.61 |
| **Density=1200** | 7 | 6.87 | 6.33 | 8.02 | 4.56 | 5 | 3.9 | 6.25 |
| **Density=1300** | 6.8 | 3.82 | 6.11 | 4.27 | 4.68 | 2.5 | 4.03 | 3.125 |
| **Density=1410** | 6.55 | 0 | 5.86 | 0 | 4.8 | 0 | 4.16 | 0 |

*Material Optimization*
*Material Parameters of Leaflets*

The 3D mechanical properties of the Magna premium valve were optimized based on the displacement from the experimental results obtained from the pulse duplicator system. In-plane displacement of the middle point of the middle edge in the FE simulation in Abaqus was matched with the experimental results. Initial and optimized material coefficients are shown in the following. The displacement results were obtained after two optimizations (initial optimization and optimization) with different initial parameters showed a proper match with the experimental results[24]. A perfect agreement in the opening and closing and the slope can be seen in Figure 7.

Initial material components of Fong model reported in the literature for the 25 mm PERIMOUNT Magna [24]:

$$\begin{bmatrix} 63.42 & 31.84 & 51.29 & 17.37 & 49.02 & 39.39 \\ & 63.74 & 46.75 & 68.38 & 63.09 & 19.22 \\ & & 62.82 & 38.51 & 60.17 & 55.50 \\ & & & 14.30 & 15.47 & 28.04 \\ & Symmetric & & & 47.30 & 13.69 \\ & & & & & 67.53 \end{bmatrix}$$

Optimized material components of Fong model found for the 25 mm PERIMOUNT Magna with 1410 kg/m³ density:

$$\begin{bmatrix} 65 & 30.519 & 46.338 & 15.81 & 46.29 & 38.554 \\ & 60 & 44.38 & 71 & 51 & 16.6 \\ & & 77 & 41.32 & 69.89 & 43 \\ & & & 12 & 11.245 & 31.95 \\ & Symmetric & & & 46.05 & 9 \\ & & & & & 67 \end{bmatrix}$$

**Table 4: Material parameter for the 3D anisotropic Fung model**

| | C | α |
|---|---|---|
| **Paramount Magna (Initial coefficients)** | 90720 | 10642 |
| **Paramount Magna (Optimized coefficients)** | 59000 | 2642 |

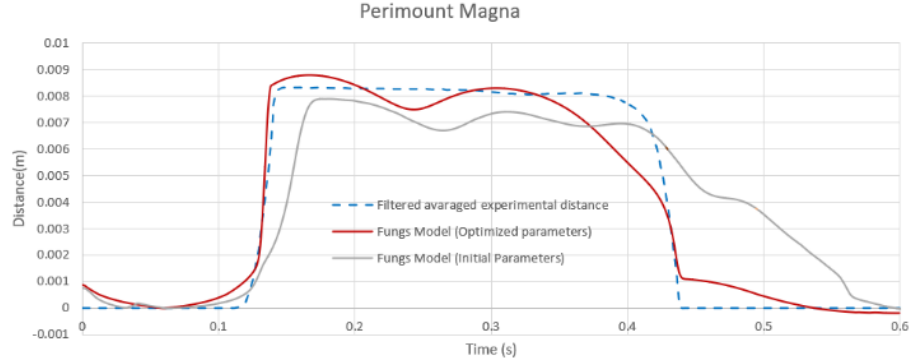

**Figure 7: Middle point displacement of the leaflet for the three valves; comparing optimized FE simulations with the experimental data**

### 4. DISCUSSION

This study improves understanding of the structural performance of surgical heart valves. Stress distribution and deformation of the BHV leaflets depend on the magnitude of density; by considering 1410 kg/m$^3$ as the reference value for density, the difference of taking the value of 1000 kg/m$^3$ into account as the density value would be remarkable. Relevant differences in stress distribution have a strict correlation with different kinematics of the leaflets. In computational models, it is essential to consider an accurate value for the density of BHV leaflets.

Computational modeling and FE analysis in particular are significantly strong tools for evaluating and optimizing the durability of BHVs through displacement and stress analysis. To obtain the mechanical behaviors of bioprosthetic aortic valve leaflets, accurate properties of tissues are essential, helping us design and manufacture novel prosthetic valves.

Calcification of the prosthesis is the main issue faced by bovine pericardial valves; these problems can result in stenosis in the valve leaflet and may ultimately cause failure. Glutaraldehyde (GA) fixation is commonly used in the fabrication of existing bioprosthetic valves [7]. Notably, pericardium anisotropy is considerably decreased by the GA fixation process [27]. There are newer approaches in the form of novel multistep procedures aimed at anti-calcification. New procedures are developed to reduce rate of calcification, including valve fixation using non-glutaraldehyde crosslinking agents; calcification prevention through prostheses treatment via trivalent metal ions or 2-alpha- amino-oleic acid; or the application of surfactants or ethanol to eliminate materials with calcification potentials [28].

In addition to efforts toward preventing calcification, studies have shown correlations between the regions of tearing in BHV leaflets, and regions of high tensile and bending stresses [29,30]. It is widely accepted that stress concentrations within leaflets can either directly accelerate tissue structural fatigue damage or stimulate calcification. BHV calcification primarily develops in regions of high mechanical stress. Accelerated stress-driven deterioration of the extracellular matrix integrity accompanied by proteolysis can potentially promote the deposition of Ca2+ on damaged collagen and elastin fibers [32]. As a result, valve hemodynamics is affected by BHV stenosis and calcification of the leaflets, which promote mechanical stress. There is also a consensus that valve designs that reduce leaflet stresses are likely to show improved performance in long-term durability [4]. Ultimately, Finite Element (FE) analysis has been used extensively for the analysis of native and prosthetic valve mechanics.

Incorporating experimentally derived material properties into FE models and simulating valve deformations under physiological conditions will ultimately improve BHV design and, consequently, durability. In this study, we overlooked variability with regards to the three leaflet material properties in simulations of three bioprostheses aimed at lowering the computational time. Furthermore, to imitate the surrounding fluid, viscous damping was applied in FE simulations. Simulating the interaction between fluid and structure could offer a simulation with a higher level of precision which we will consider in future studies. Also, adding more samples could increase the accuracy of our experiment. Thus, more robust findings could be achieved from experiments involving more patches.

### 5. CONCLUSION

In summary, a rigorous experimental evaluation was carried out to obtain the accurate density of the bovine pericardium patch. The impact of different densities on SHV leaflet deformation and stress distribution was taken into account. It was found that the bovine pericardium patch density is higher than the amount used in simulations. SHV leaflet stress and strain distributions depended on the magnitude of density. It was also found that density difference in simulations induced higher deformation and high-stress regions within the commissures in the acceleration, while inducing localized low-stress regions within the belly of the TAV leaflets during the deceleration phase of the cardiac cycle. Through dynamic simulations of a 25-mm PERIMOUNT Magna aortic valve, it was shown that the maximum principal stress was significantly higher in the acceleration phase of cardiac cycle than the deceleration phase in all cases.

**FUNDING STATEMENT**
This work was supported by the University of Denver (grant number: 84993-142235).

**CONFLICT OF INTEREST**
The authors have no conflict of interest to declare.